\documentclass[twocolumn]{aastex631}
\usepackage{multirow}
\usepackage{dcolumn}

\newcommand{\gr}{$\gamma$-ray}

\newcommand{\fermi}{{\it Fermi}}

\shorttitle{$\gamma$-ray emissions of PSR~J1717$+$4308A and other globular-cluster millisecond pulsars}
\shortauthors{Zhang et al.}
\graphicspath{{./}{figures/}}

\begin{document}

\title{Likely detection of $\gamma$-ray pulsations of PSR~J1717$+$4308A in NGC~6341 and implication of the $\gamma$-ray millisecond pulsars in globular clusters}

\author{Pengfei Zhang}
\affiliation{Department of Astronomy, School of Physics and Astronomy, Key Laboratory of Astroparticle Physics of Yunnan Province, Yunnan University, Kunming 650091, People's Republic of China; zhangpengfei@ynu.edu.cn; wangzx20@ynu.edu.cn}

\author[0000-0002-5818-0732]{Yi Xing}
\affiliation{Key Laboratory for Research in Galaxies and Cosmology, Shanghai Astronomical Observatory, Chinese Academy of Sciences, 80 Nandan Road, Shanghai 200030, People's Republic of China}

\author[0000-0003-1984-3852]{Zhongxiang Wang}
\affiliation{Department of Astronomy, School of Physics and Astronomy, Key Laboratory of Astroparticle Physics of Yunnan Province, Yunnan University, Kunming 650091, People's Republic of China; zhangpengfei@ynu.edu.cn; wangzx20@ynu.edu.cn}
\affiliation{Key Laboratory for Research in Galaxies and Cosmology, Shanghai Astronomical Observatory, Chinese Academy of Sciences, 80 Nandan Road, Shanghai 200030, People's Republic of China}

\author{Wei Wu}
\affiliation{Department of Astronomy, School of Physics and Astronomy, Key Laboratory of Astroparticle Physics of Yunnan Province, Yunnan University, Kunming 650091, People's Republic of China; zhangpengfei@ynu.edu.cn; wangzx20@ynu.edu.cn}

\author{Zhangyi Chen}
\affiliation{Department of Astronomy, School of Physics and Astronomy, Key Laboratory of Astroparticle Physics of Yunnan Province, Yunnan University, Kunming 650091, People's Republic of China; zhangpengfei@ynu.edu.cn; wangzx20@ynu.edu.cn}

\begin{abstract}
We report our analysis results for the globular cluster (GC) NGC~6341 (M92), 
as a millisecond pulsar (MSP) J1717$+$4308A has recently been reported found 
in this GC. The data used are from the Large Area Telescope onboard the 
{\it Fermi Gamma-ray Space Telescope (Fermi)}.  We detect $\gamma$-ray 
pulsations of the MSP at a $4.4\sigma$ confidence level (the corresponding 
weighted H-test value is $\sim$28.4). This MSP, the fourth $\gamma$-ray pulsar 
found in a GC, does not have significant off-pulse emission and has 
$\gamma$-ray luminosity and efficiency $1.3\times10^{34}$\,erg\,s$^{-1}$ 
and 1.7\% respectively. In order to have a clear view on the properties of 
the known GC $\gamma$-ray MSPs, we re-analyze the \fermi\ LAT data for 
the other three ones.  These four MSPs share the properties of either having 
high $\dot{E}$  ($\sim 10^{36}$\,erg\,s$^{-1}$) or being in the GCs that 
contain only limited numbers of known MSPs. In addition, we find that
PSRs~J1823$-$3021A and B1821$-$24, in NGC~6624 and NGC~6626 respectively, have
detectable off-pulse $\gamma$-ray emission and PSR J1835$-$3259B in NGC~6652
does not. Using the obtained off-pulse spectra or spectral upper limits, 
we constrain the numbers of other MSPs in
the four GCs. The results are consistent with the numbers of the radio pulsars
reported in them. While at least in NGC~6624 and NGC~6626, the contribution
of other MSPs to their observed $\gamma$-ray emission can not be ignored,
our study indicates that the presence of a bright MSP could be the dominant
factor for whether a GC is detectable at $\gamma$-rays or not.
\end{abstract}


\section{Introduction}
\label{Intro}
Thanks to the successful launch of the Large Area Telescope (LAT) onboard
{\it the Fermi Gamma-ray Space Telescope} \citep[{\it Fermi};][]{Atwood2009},
it has been found that the Galactic globular clusters (GCs) can have 
significant \gr\ emissions. The GC 47 Tucanae was the first one reported
with the \gr\ emission \citep{Abdo2009}, then followed with 
Terzan~5 \citep{Kong2010}. Up to now, \gr\ emissions from approximately 
35 GCs have been reported \citep{Tam2011a,ED2012,Zhou2015,Zhang2016,Tam+2016,Abdollahi2020,Son+2021,wu+22,yua+2022,4fgl-dr3}.
It is generally recognized that millisecond pulsars (MSPs) contained in the GCs
are the sources of the emissions, which is supported by the facts
that there are 257 MSPs (spin period $P\leq 30$\,ms) found in 36 GCs (based on the GC pulsar table\footnote{https://naic.edu/$\sim$pfreire/GCpsr.html})
and \gr~pulsations of three of them have been detected in the LAT data.
\begin{figure*}
\centering
\includegraphics[angle=0,scale=0.45]{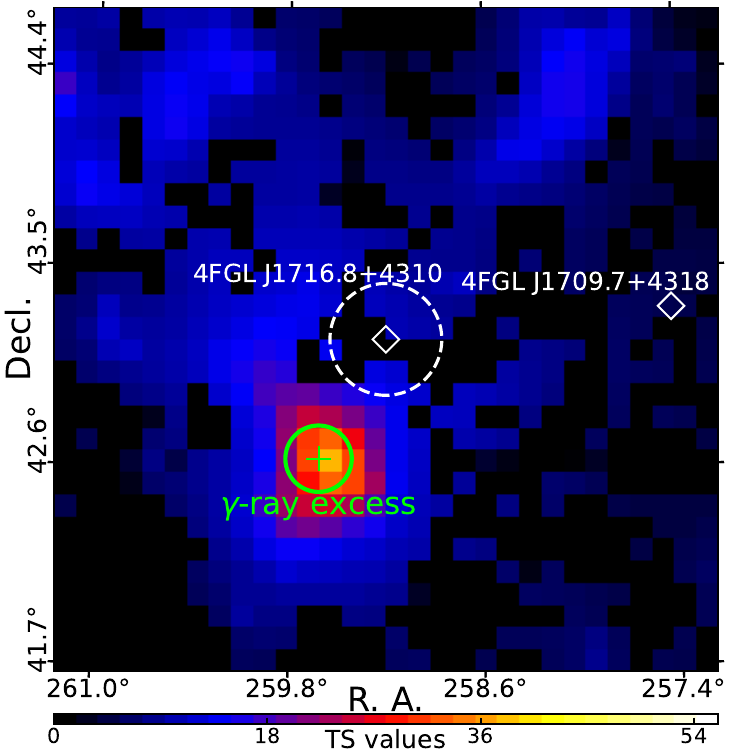}
\includegraphics[angle=0,scale=0.45]{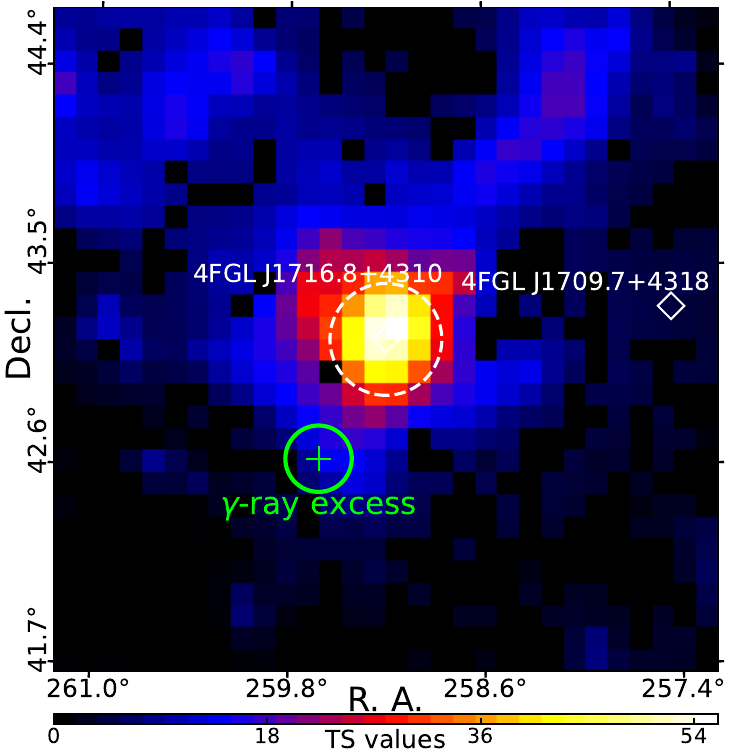}
\caption{TS maps of a $3^{\circ}\times3^{\circ}$ region centered at 
	4FGL~J1716.8$+$4310 in 0.1--500\,GeV, where
	the {\it left} was based on 4FGL-DR3, showing a \gr\
	source located south-east of the center, and the {\it right} was
	calculated when the new source was considered and removed 
	(while 4FGL~J1716.8$+$4310 was kept). The position of
	the new source and its $3\sigma$ 
	uncertainty are marked by the green plus and circle respectively,
	and the white dashed circle indicates the tidal radius of NGC~6341.
	The pixel scale of the both TS maps is $0\fdg1$\,pixel$^{-1}$. }
\label{fig:tsmap}
\end{figure*}

The first reported \gr--pulsations case was PSR~J1823$-$3021A in NGC~6624,
which had a significance of $\sim7\sigma$ \citep{fre+11}.
Then in NGC~6626 (M28), significances of $\sim4.3\sigma$ and $5.4\sigma$ 
\gr~pulsations were respectively reported for PSR~B1821$-$24 by \citet{wu+13} 
and \citet{joh+13}.
Very recently, \citet*{zxw22} reported the detection of \gr~pulsations
(at a $5.4\sigma$ significance level) of
a newly discovered MSP J1835$-$3259B in NGC 6652 \citep{gau+2022}.
The properties of these three MSPs are summarized in Table~\ref{tab:prop}.
They share the similarities of having relatively high spin-down energies
$\dot{E}$, their \gr\ luminosities being among the brightest ones of all
known \gr\ MSPs \citep{zxw22}, and they probably contribute a dominant 
fraction to the \gr\ emission of each of the respective host GCs.

More such \gr\ MSPs may be found in GCs, provided accurate radio 
ephemerides are available. We note that
\citet{pan+20} reported the discovery of PSR~J1717$+$4308A in the GC 
NGC~6341 (M92) from the observations with the Five-hundred-meter Aperture 
Spherical radio Telescope (FAST).
The MSP, having $P=3.16$~ms,
is in an eclipsing binary system with $\sim$0.2-d orbital period.
While this MSP is the only pulsar thus-far found in the GC, 
the \gr\ emission from the GC has been detected and in the 
Data Release 3 (DR3) of the fourth \emph{Fermi}-LAT source catalog (4FGL-DR3;
\citealt{4fgl-dr3}), the \gr\ source is named 4FGL J1716.8$+$4310.
Given these and the detailed ephemeris available for 
PSR~J1717$+$4308A \citep{pan+20,pan+21}, we have carried out analysis of 
the \fermi-LAT data for finding the \gr~pulsations of the MSP.
We report the result, a $4.4\sigma$ significance detection of 
the \gr~pulsations, in this paper.

\begin{table}
\begin{center}
\caption{Properties of GC MSPs with \gr\ pulsations detected}
\begin{tabular}{lccccc}
\hline\hline
	Name & $P$  & $\dot{P}/10^{-20}$ & $\dot{E}/10^{34}$ & $L_{\gamma}/10^{34}$ \\
	& (ms) & (s~s$^{-1}$) & (erg~s$^{-1}$) & (erg~s$^{-1}$) \\
\hline
J1823$-$3021A & 5.44 & 338 & 83  & $7.0\pm1.8^{\rm a}$ \\
B1821$-$24   & 3.05 & 161 & 220 & 4.0$\pm1.0^{\rm b}$ \\
J1835$-$3259B & 1.83 & 6.65 & 43  & $5.04\pm0.44^{\rm c}$ \\
J1717$+$4308A & 3.16 & 6.11 & 7.7  & $1.30\pm0.17$ \\
\hline\hline
\end{tabular}
\label{tab:prop}
\end{center}
	{\textbf{Notes.} References for $L_{\gamma}$: $^{\rm a}$ \citet{2pc13}, $^{\rm b}$ \citet{joh+13}, and $^{\rm c}$ \citet{zxw22}.}
\end{table}

In addition, considering that \gr\ emissions of the GCs arise from their
MSPs, the high-energy studies provide a probe into the
population of MSPs in the GCs besides the regular radio compaigns. The numbers 
of the MSPs in the GCs have been estimated by 
comparing the isotropic \gr~luminosity of a GC with the average \gr\ luminosity
of the MSPs \citep{Abdo2009,abd+10,hui+11,lcb18,dcn19,zzy+20,Son+2021}.
Recently, \citet{wu+22} have added the \gr\ spectral information of the
MSPs into the estimation, which presumably have provided better constraints
on the numbers of MSPs in more than 20 GCs. The numbers estimated in
this latter work indicate that for approximately 10 GCs, only one MSP may be
required in them to explain the observed \gr\ emissions. Combined with
the likely dominant contributions of the detected three GC MSPs to the \gr\ 
emissions of their respective host GCs \citep{fre+11,wu+13,joh+13,zxw22},
an interesting question may be raised as if the \gr\ emissions of most GCs
are similarly due to the presence of one bright \gr\ MSP (i.e., not reflecting
the MSP numbers). This possibility may explain why some of GCs, while
with high encounter rates, do not have detectable \gr\ emissions
(e.g., see \citealt{wu+22}); it would be rather because they do not
contain one sufficiently bright MSP. To investigate this possibility, we 
also conducted time analysis of the LAT data for the three known GC \gr\ MSPs.
For each of them, the off-pulse emission was obtained, so as to set a
constraint on the number of other MSPs in each of the host GCs (see, e.g.,
\citealt{joh+13}).
This part of analysis and the results are also reported in this paper.

Below we first describe the data analysis for PSR~J1717$+$4308A in
NGC~6341 (M92) and report the results in Section~\ref{sec:j17}. We 
describe the analysis for the three GCs that contain a known \gr\ MSP
and present
the results in Section~\ref{sec:three}. In Section~\ref{sec:dis},
we discuss the results and implications for MSPs in GCs and provide
a summary.

\begin{table}
\begin{center}
\caption{Likelihood analysis results for NGC~6341}
\begin{tabular}{ccccc}
\hline\hline
Model & \multicolumn{3}{c}{Parameter values} \\
\hline
LP& $\alpha$ & $\beta$ & $E_b$ (GeV) & TS \\
 & 2.03(31)$^\ast$ & 0.45(24)$^\ast$ & 1.17$^\ast$ & --\\
	& 1.88(13) & 0.71(11) & 1.17(16) & 63.7 \\
\hline
PLEC & $\Gamma$ & $b^\dagger$ & $E_c$ (GeV) \\
 & 1.88(17) & 0.67 &3.00(52) & 63.9\\
\hline
\end{tabular}
\label{tab:par}
\end{center}
    {$^\ast$Parameter values of the LP model given in 4FGL-DR3, among which
	$E_b$ does not have an uncertainty. $^\dagger$ Value was fixed at 2/3 in our analysis.}
\end{table}

\section{Data analysis and results for PSR~J1717$+$4308A}
\label{sec:j17}

\subsection{LAT data selection and best-fit model}
\label{sec:model}

We selected 14 year \emph{Fermi}-LAT Pass 8 \emph{Front+Back} events 
(evclass = 128 and evtype = 3) in a time range of from 2008 August 4 
16:29:16.8 (MJD~54682.69) to 2022 September 25 22:22:29.0 (MJD~59847.93) 
in the energy range of 0.1--500 GeV.  
The region of interest (RoI) was $20^\circ\times20^\circ$ with the center
at the position of 4FGL J1716.8$+$4310
(R.~A.~=~$\rm17^h16^m48^s.79$, Decl.~=~$+43^{\circ}10'22\farcs78$).
The events with zenith angles $<90^{\circ}$ were kept to avoid the limb 
contamination from the Earth. We used the expression of 
``DATA\_QUAL $>$ 0 \&\& LAT\_CONFIG == 1" to filter out the events with 
quality flags of ``bad" and to save high-quality events in the good time 
intervals. The instrument response function of “P8R3\_SOURCE\_V3” and 
the software package of Fermitools–2.0.19 were used in the analysis.

Based on 4FGL-DR3, we constructed a model file using the script 
make4FGLxml.py\footnote{https://fermi.gsfc.nasa.gov/ssc/data/analysis/user/}.
The model file contained the spectral forms of the catalog sources around 
4FGL~J1716.8$+$4310 and the respective parameters.
All spectral parameters of the sources within $5^{\circ}$ of the target
were set free, and for the sources within $5^{\circ}-10^{\circ}$ of the target 
and those 
$10^{\circ}$ outside but identified as variables, their normalization
parameters were set free. In addition, the normalizations of the two diffuse 
emission components (Galactic and extragalactic) were also set free. 
All other parameters of the sources in the RoI were fixed at the values 
provided in 4FGL-DR3.
The target is described with a log-parabola (LP) model, 
$dN/dE=N_0(E/E_b)^{-[\alpha+\beta\log(E/E_b)]}$, in 4FGL-DR3. The catalog
parameter values are given in Table~\ref{tab:par}. 
\begin{figure}
\centering
\includegraphics[angle=0,scale=0.48]{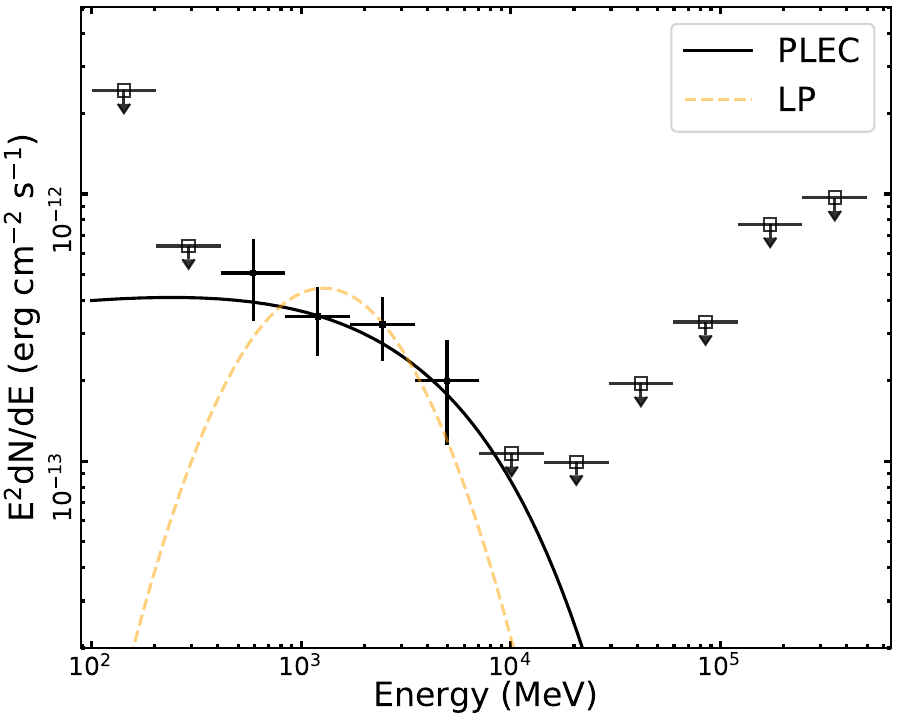}
\caption{\gr~spectrum in 0.1--500 GeV obtained for 4FGL~J1716.8$+$4310.
The best-fit LP and PLEC model spectra are shown as the yellow dashed and black 
	solid lines respectively. }
\label{fig:spec}
\end{figure}

The binned maximum likelihood analysis was performed to the whole data.
In the results we found a new \gr\ source that was not previously reported
or listed in the \fermi-LAT catalogs. As shown in the Test Statistic (TS) map
(Figure~\ref{fig:tsmap}) we calculated for the source region, this source 
is located south-east of the target and had a TS value of $\sim$40.
Using tool \emph{gtfindsrc}, its position was determined to be
R.~A.~=~17$^h$18$^m$29$^s$.79, Decl.~=~+42$^{\circ}$38$'$37$''$.38 (J2000.0).
We added a point source with a spectral form of power-law (PL), 
$dN/dE=N_0(E/E_0)^{-\Gamma}$, at its position in our model file.
Performing the likelihood analysis again, the source could be well fitted
and removed in the TS map (Figure~\ref{fig:tsmap}). Detailed analyses and
results for this new source will be reported elsewhere (Zhang et al. in 
preparation).
The best-fit parameter values for 4FGL~J1716.8$+$4310, after adding 
the new source in our
model file, are given in Table~\ref{tab:par}. Comparing them with the
catalog values, there are small differences but within
the uncertainties.
\begin{figure*}
\centering
\includegraphics[angle=0,scale=0.56]{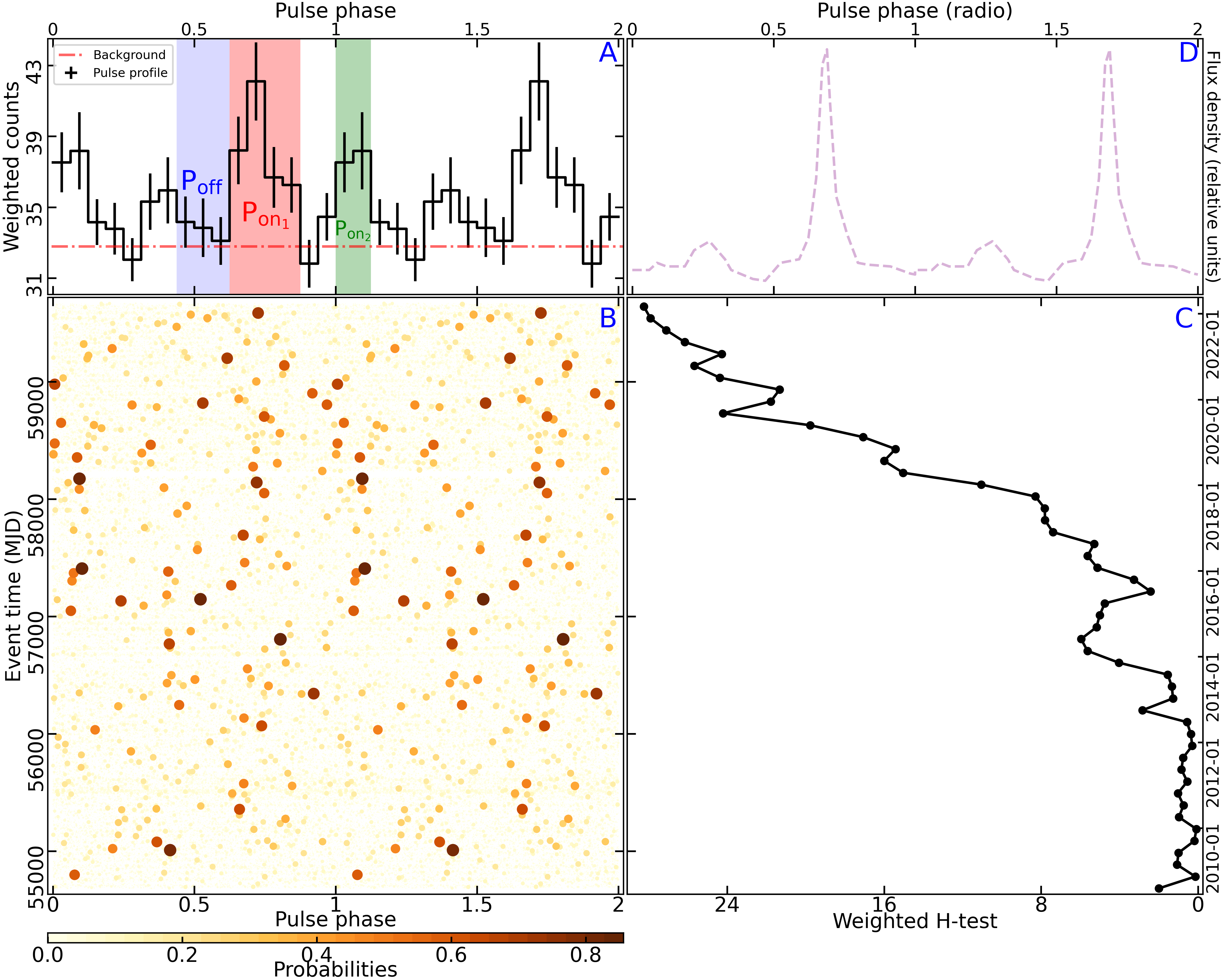}
\caption{Timing analysis results for PSR~J1717$+$4308A in 0.1-500 GeV.
        Panel A: weighted \gr~pulse profile. The on-pulse
	phase ranges, $\rm P_{on_1}$ and $\rm P_{on_2}$, and off-pulse range
	$\rm P_{off}$ are marked as
        pink, green, and blue shaded regions respectively. The background 
	counts is shown as the pink dash-dotted line.
        Panel B: two-dimensional phaseogram, where the color bar indicates
	the probabilities of \gr\ photons originating from NGC~6341
	(the largest probability value is 85.6\%).
	Panel C: weighted cumulative H-test curve over the whole
	time period of the \fermi-LAT data.
	Panel D: schematic radio pulse profile of PSR~J1717$+$4308A 
	drawn based on that reported in \citet{pan+20} for comparison.
        }
\label{fig:fold}
\end{figure*}

As the emission should have an MSP origin, we also considered 
the typical pulsar model, a PL with an exponential cutoff 
(PLEC; \citealt{2pc13}),
$dN/dE=N_0(E/E_0)^{-\Gamma}\exp[-(E/E_c)^b]$. Replacing with
this PLEC
in our model file for 4FGL~J1716.8$+$4310, the binned likelihood analysis
was performed, in which the parameter $b$ was fixed at 2/3 \citep{2pc13}.
The resulting best-fit parameters are given in Table~\ref{tab:par}.
The TS value from this PLEC model is nearly equal to that from the LP model,
and thus this model can also be used for the emission of the source.
The average photon flux obtained was 
$2.43\pm0.31\times10^{-9}$\,photon\,cm$^{-2}$\,s$^{-1}$ in 0.1$-$500\,GeV,
giving an integrated energy flux of 
$\rm 1.51\pm0.19\times10^{-12}~erg~cm^{-2}~s^{-1}$. Using a distance
of 8.50$\pm0.07$\,kpc for NGC~6341\footnote{https://people.smp.uq.edu.au/HolgerBaumgardt/globular/fits/ngc6341.html},
the \gr~luminosity $L_{\gamma}\simeq 1.30\pm0.17\times10^{34}$\,erg\,s$^{-1}$
(assuming isotropic emission).

\subsection{Spectrum extraction}

We extracted a spectrum of 4FGL~J1716.8$+$4310 from the whole LAT data.  
The energy range of 0.1--500\,GeV was divided into 12 equal logarithmically 
spaced energy bins, and the likelihood analysis was performed to the data of
each bins.
In this analysis, only the normalizations of the sources within 5$^{\circ}$
of the target, which included the new nearby source we have found,
were set as free parameters, while all other parameters were fixed at their 
values obtained in the above likelihood analysis.
The obtained spectrum is shown in Figure~\ref{fig:spec}, for which we kept 
the spectral data points that have their values 2 times greater than their
uncertainties and otherwise showed 
95\% flux upper limits.  Comparing the best-fit LP and 
PLEC model to the spectrum, the latter model appears to be able to fit 
the spectrum as well.
Given this and the nearly equal TS values from the two models, we adopted
the PLEC model in the following data analysis.
\begin{figure*}
\centering
\includegraphics[angle=0,scale=0.45]{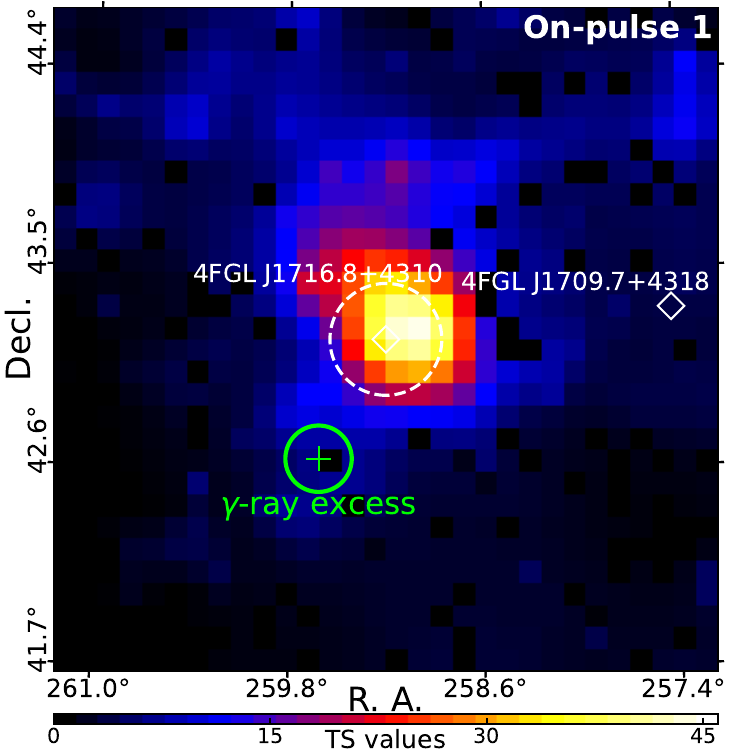}
\includegraphics[angle=0,scale=0.45]{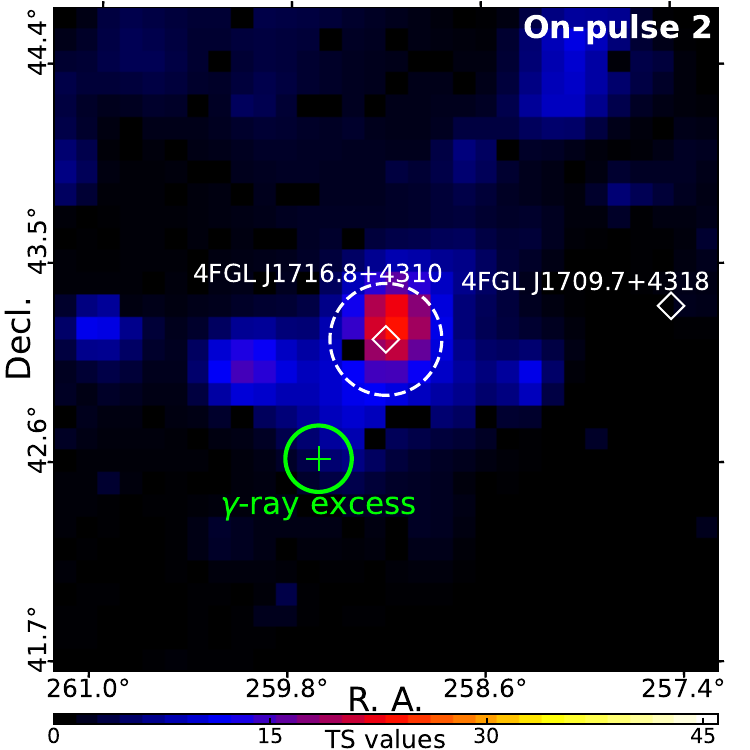}
\includegraphics[angle=0,scale=0.45]{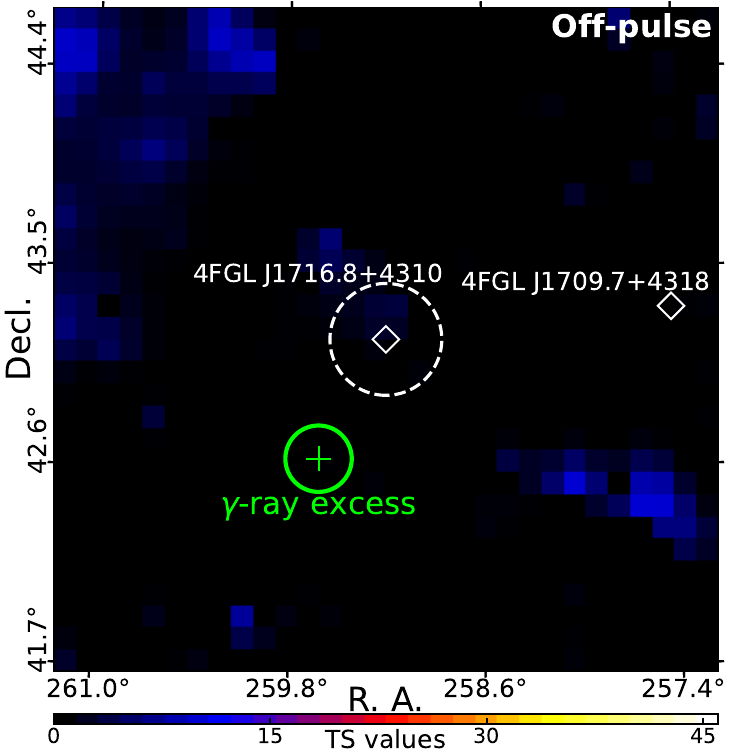}
\caption{TS maps in 0.1--500\,GeV calculated from the data in two on-pulse 
	({\it left} and {\it middle} panels)
	and one off-pulse ({\it right} panel) phase ranges. 
	The tidal radius region of NGC~6341
	is marked as the white dashed circle, and the new \gr\ source is also
	indicated as that in Figure~\ref{fig:tsmap}. The pixel scale of 
	the TS maps is 0\fdg1\,pixel$^{-1}$.}
\label{fig:phase-tsmap}
\end{figure*}

\subsection{Timing Analysis}

Since the \gr~emission from NGC 6341 is weak (TS$\simeq$64) and there are 
nearby sources with comparable brightnesses, we followed \citet{abdo+10} and
selected the events within an aperture radius of 
$max[6.68-1.76\log_{10}(E_{MeV}), 1.3]^{\circ}\simeq 3\fdg16$ in 
0.1-500\,GeV so as to maximize the signal-to-noise ratio.
The selected \gr\ photons were assigned with the weights, which were 
the probabilities of originating from NGC~6341 given by tool 
\emph{gtsrcprob}.
The weighted photons were then phase-folded using the ephemeris 
for PSR~J1717+4308A provided in \citet{pan+20,pan+21} by employing 
Tempo2 \citep{hem06} with \emph{Fermi} plug-in \citep{rkp+11}. 
The folded pulse profile in 16 phase bins is shown in Panel~A of
Figure~\ref{fig:fold}, for which the counts and uncertainties of
the phase bins were calculated using the method described in \citet{2pc13}.
The two-dimensional phaseogram is shown in Panel~B of Figures~\ref{fig:fold},
and the largest probability value of the photons was 85.6\%.

Following the method described by \citet{k11}, H-test statistic was calculated
using the probabilities as the weights.
The curve of the cumulative H-test value over the whole \fermi-LAT data
time period is shown in Panel~C of Figure~\ref{fig:fold}.
The maximum H-test value was $\sim$28.4, corresponding to
a $p$-value of $1.17\times10^{-5}$ ($\simeq 4.4\sigma$).
It can be noted that the H-test curve is flat and close to zero before year 
2014.  We suspect that since the ephemeris for the MSP was derived from
the FAST observations conducted from 2017 October and the MSP is in 
a redback binary \citep{pan+20}, the type showing  
high and random orbital variability (see, e.g., \citealt{rid+16}),
the phases of the photons before year 2014 could not be accurately given.


\subsection{Phase-resolved Analysis}

Given the folded pulse profile, we divided it
into the on-pulse and off-pulse regions. For the former, two peak 
phase regions, $P_{\rm on_1}$ and $P_{\rm on_2}$,
were chosen as in phase ranges of 0.62--0.88 and 1.00--1.12 respectively.
The latter, $P_{\rm off}$, was chosen in a phase range of 0.44--0.62.
The likelihood analysis was performed to the data in the three phase ranges
respectively, in which the PLEC model was used.
In $\rm P_{on_1}$, the resulting parameters were $\Gamma=2.05\pm0.28$ and 
$E_c=3.00\pm0.66$ ($b=0.67$ was fixed), consistent with those from
the whole data (Table~\ref{tab:par}); the photon flux was
$6.6\pm1.3 \times10^{-9}$ photon\,cm$^{-2}$\,s$^{-1}$ with a TS value of 
$\sim$49.
In $\rm P_{on_2}$, the TS value was only $\sim$25 (the photon flux was
$4.6\pm 1.4\times10^{-9}$\,photon\,cm$^{-2}$\,s$^{-1}$), 
and in $\rm P_{off}$,  TS$\sim$1 was obtained.
The nearly zero TS value in $\rm P_{off}$ is expected 
as the counts in the phase bins
are close to the background value (=32.8; cf., Figure~\ref{fig:fold});
the latter was estimated
from diffuse sources and neighbor point sources following \citet{2pc13}.

We calculated the corresponding TS maps for the three phase intervals, 
which are shown in Figure~\ref{fig:phase-tsmap}. The TS maps support
the timing analysis results. A simple direct comparison is that while 
the phase intervals of 
$\rm P_{on_2}$ and $\rm P_{off}$ are 0.12 and 0.18,
the TS values are 25 and 1 respectively. The \gr\ emission from NGC~6341
thus seems to originate, within current measurement precision, 
exclusively from PSR~J1717+4308A. In the future as more data are collected,
which improves the sensitivity, it would become clear if there is
significant off-pulse emission from this GC (see the case of NGC~6624 below
in Section~\ref{subsec:pra}).  

\section{Data analysis for three known GC \gr\ MSPs}
\label{sec:three}

In order to separate a known pulsar's emission from that of the host GC,
we carried out the
similar data analysis for NGC~6624 and NGC~6626 as for NGC~6341. 
The analysis results 
for NGC~6652 were taken from that reported in \citet{zxw22}.
\begin{figure*}
\centering
\includegraphics[angle=0,scale=0.35]{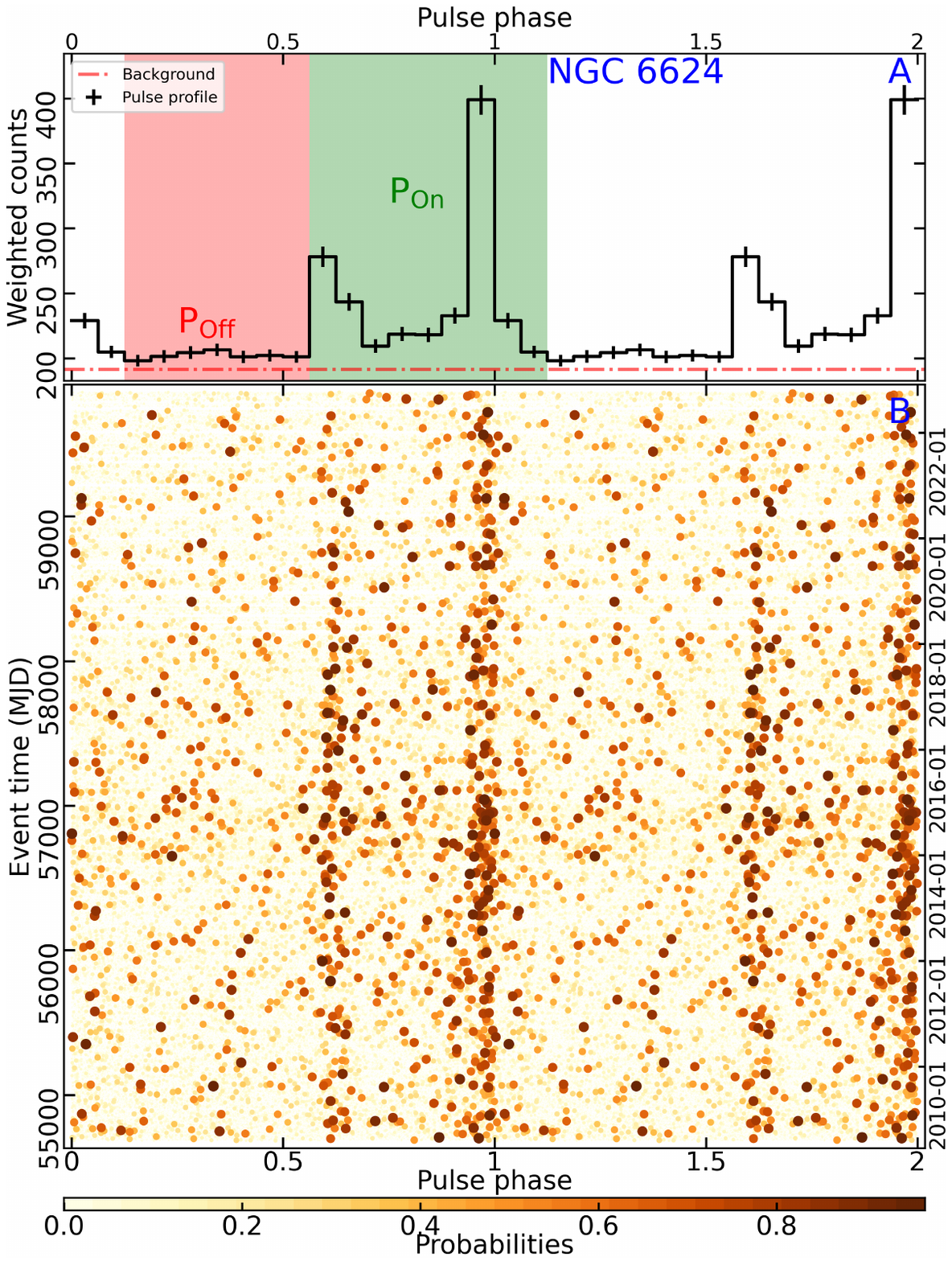}
\includegraphics[angle=0,scale=0.35]{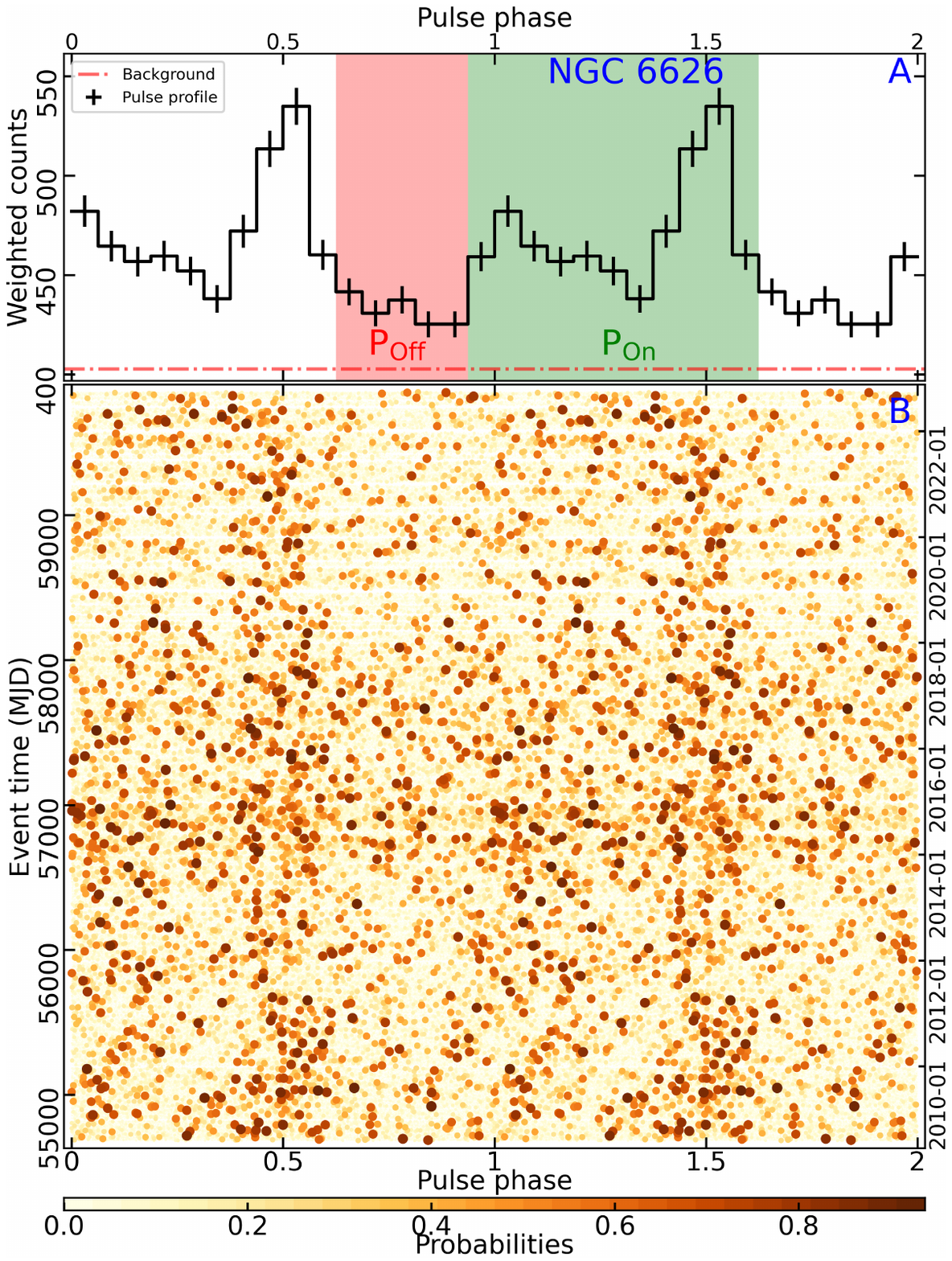}
\caption{Pulse profiles and two-dimensional phaseograms obtained for 
	PSRs~J1823$-$3021A ({\it left}) and B1821$-$24 ({\it right}) 
	respectively. Based on the pulse profiles, the on-pulse
	($\rm P_{on}$) and off-pulse ($\rm P_{off}$) phase ranges are defined.}
\label{fig:fold2426}
\end{figure*}

For each of the two target GCs, the LAT data of similar time periods in
the same energy range (0.1--500\,GeV) were selected. Likelihood analysis 
was performed to the whole data to determine the best-fit parameters,
in which a PLEC model was used. We then updated the model files for each
of them.

\subsection{Timing Analysis}

To construct the pulse profiles of PSRs J1823$-$3021A \citep{big+94} 
and B1821$-$24 (\citealt{lyn+87}; in NGC~6624 and NGC~6626 respectively), 
we selected photons within an 
aperture radius of 6$^\circ$ centered at each of the positions given 
in \citet{4fgl-dr3}. The ephemerides of the two MSPs are provided by the 
Fermi Science Support Center\footnote{https://fermi.gsfc.nasa.gov/ssc/data/access/lat/ephems/}.
The weights of the photons were the probabilities of
originating from a target calculated using tool \emph{gtsrcprob}, and
the pulse phases of them were assigned by employing Tempo2 with the \fermi\ 
plug-in.
Given the $\sim$14-yr long time period of the data, the ephemerides were 
updated from running Tempo2 by us.
The resulting folded pulse profiles and two-dimensional phaseograms for
PSRs J1823$-$3021A and B1821$-$24 are shown in
Figure~\ref{fig:fold2426}.
The weighted H-test values were $\sim$1240 and $\sim$213 for J1823$-$3021A 
and B1821$-$24 respectively.

\subsection{Phase-resolved analysis}
\label{subsec:pra}

Based on the pulse profiles we obtained, which are
very similar to those shown in \citet{fre+11} and \citet{joh+13},
we defined the on-pulse ($\rm P_{on}$) and off-pulse ($\rm P_{off}$) 
phase ranges:
$\rm P_{on}\simeq$0.56--1.13 and 0.94--1.63, $\rm P_{off}\simeq$0.13--0.56 
and 0.63--0.94 respectively for the MSPs in NGC~6624 and NGC~6626.
Binned likelihood analysis was performed to the data sets respectively,
where the PLEC model was used.
The results are given in Table~\ref{tab:phar}. 
Significant off-pulse emissions from each of the GCs were detected,
while we note that in the initial discovery of \gr\ pulsations of
PSR~J1823$-$3021A that used 2-yr \fermi\ LAT data, no off-pulse emission
was detected from NGC~6624.
\begin{table}
\begin{center}
\caption{Pulse-phase resolved analysis results for NGC~6624 and NGC~6626}
\begin{tabular}{ccccc}
\hline\hline
	GC & $F_{\gamma}/10^{-11}$  & $L_{\gamma}/10^{34}$ & $\Gamma$ & $E_c$ \\
	& (erg\,cm$^{-2}$\,s$^{-1}$) & (egr\,s$^{-1}$) & & (GeV) \\
\hline
NGC 6624 \\
	$\rm P_{on}$ & 2.2(0.3) & 16.8(3.0) & 1.9(0.4) & 3.0(0.3) \\
	$\rm P_{off}$ & 0.5(0.1) & 4.2(0.9) & 1.5(0.2) & 3.0(0.1) \\
NGC 6626 \\
	$\rm P_{on}$ & 2.6(0.2) & 9.0(0.7) & 1.4(0.1) & 1.0(0.2) \\
	$\rm P_{off}$ & 1.1(0.4) & 3.8(1.4) & 1.4(0.5)& 1.3(1.1) \\
\hline\hline
\end{tabular}
\label{tab:phar}
\end{center}
    {\textbf{Notes.} NGC 6624 at $8.02\pm0.11$ kpc and NGC 6626 at $5.37\pm0.10$ kpc \citep{bv21} are used.}
\end{table}

\begin{figure*}
\centering
\includegraphics[angle=0,scale=0.45]{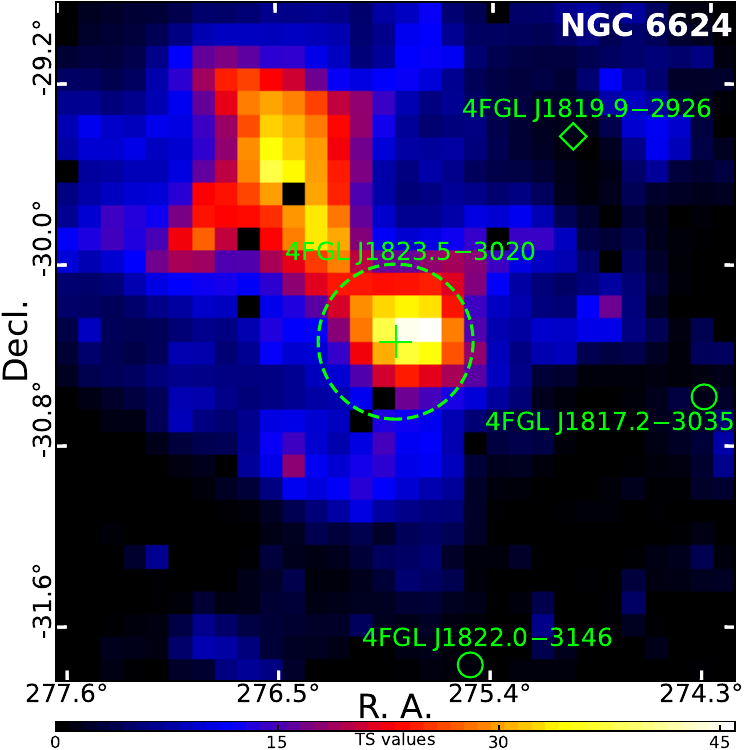}
\includegraphics[angle=0,scale=0.45]{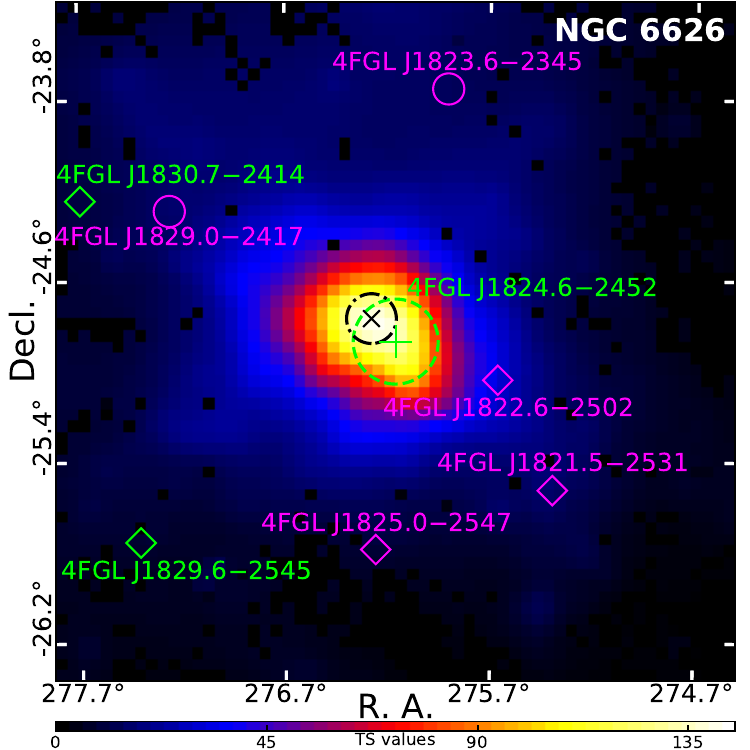}
\includegraphics[angle=0,scale=0.45]{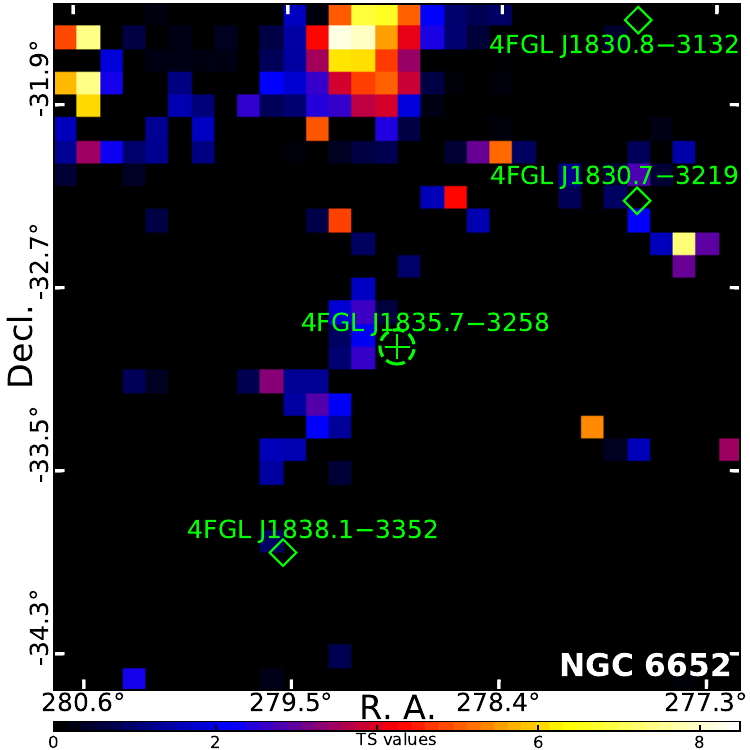}
	\caption{TS maps in 0.1--500\,GeV calculated from the off-pulse phase 
	ranges for NGC~6624 (PSR~J1823$-$3021A; {\it left}), NGC~6626 
	(PSR~B1821$-$24; {\it middle}), 
	and NGC~6652 (PSR~J1835$-$3259B; {\it right}). In each panel,
	the center and the region
	of the tidal radius for the GC are marked by the green
	plus and dashed
	circle respectively. The pixel scale of the maps is 
	0\fdg1\,pixel$^{-1}$. For the off-pulse emission of NGC~6626, its
	position and 2$\sigma$ uncertainty are marked by the black cross 
	and dash-dotted circle respectively.}
\label{fig:offtsmaps}
\end{figure*}
\begin{figure*}
\centering
\includegraphics[angle=0,scale=0.48]{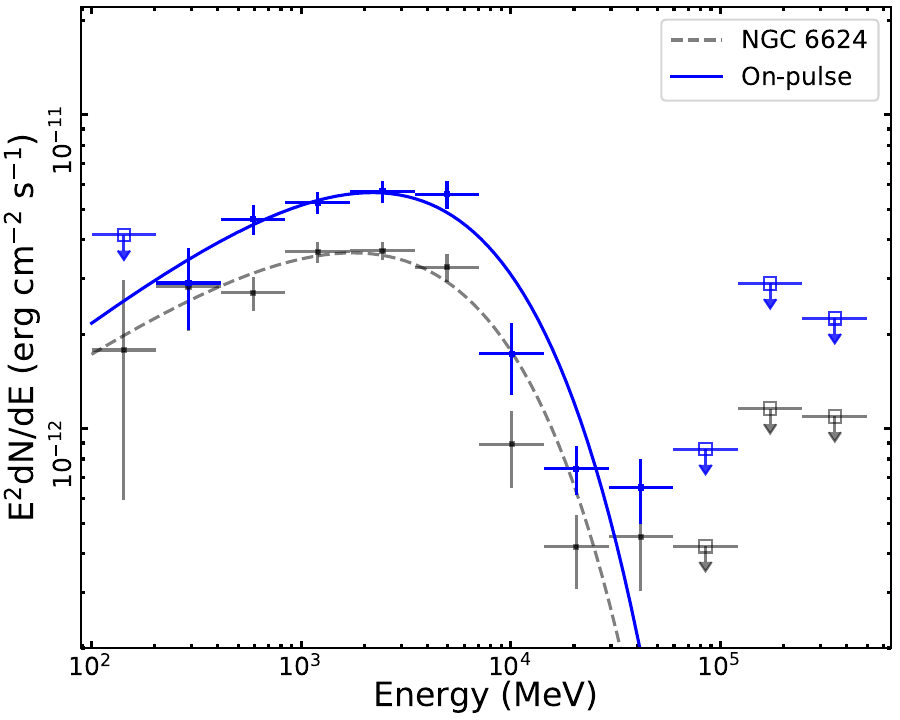}
\includegraphics[angle=0,scale=0.48]{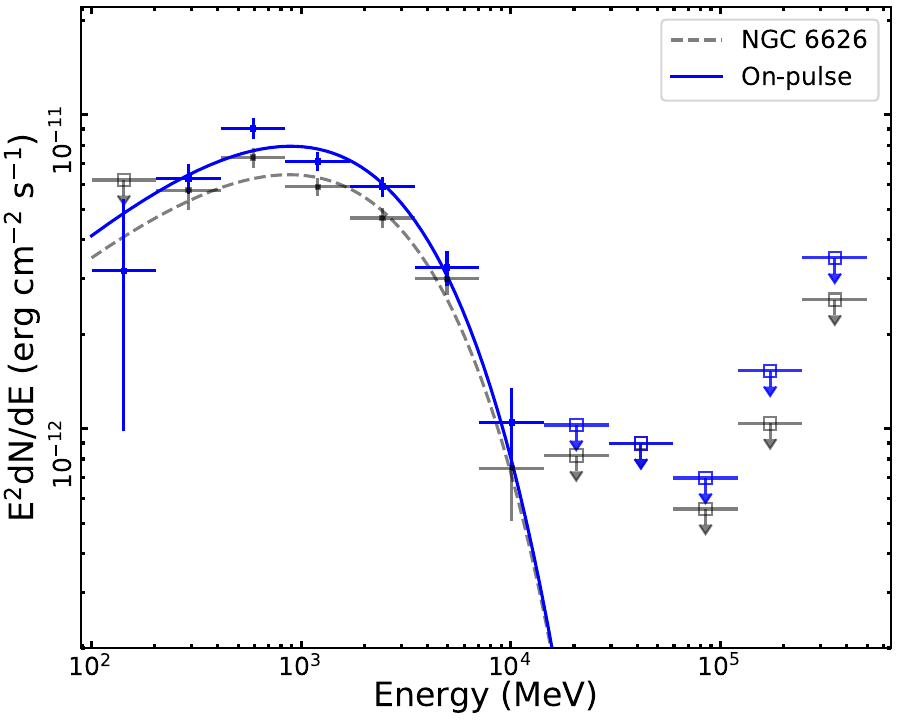}
	\caption{\gr~spectra in 0.1--500 GeV obtained for NGC~6624 ({\it left}
	panel) and NGC~6626 ({\it right} panel) and those during the on-pulse 
	phase range of PSRs J1823$-$3021A and B1821$-$24 respectively.
	In each panel, the best-fit PLEC model spectra are shown as the gray
	dashed and blue solid lines respectively. }
\label{fig:spec2}
\end{figure*}
We calculated the 0.1--500\,GeV TS maps to examine the appearances of 
the off-pulse emissions, which are shown in Figure~\ref{fig:offtsmaps}.
Within NGC~6624, TS$\sim$50 emission is present, while there is also some
residual
emission (TS$\sim$30) north-east of the GC. The best-fit parameters determined
for the off-pulse emission are consistent with those obtained for the on-pulse
one, although with large uncertainties. As already illustrated
by \citet{joh+13} in their analysis for NGC~6626, we also found
off-pulse emission similarly
appearing off the GC (Figure~\ref{fig:offtsmaps}). 
Using tool \emph{gtfindsrc}, we determined its position:
R.~A.~=~18$^h$25$^m$07$^s$.0, Decl.~=~$-$24$^{\circ}$46$'$05$''$.9 (J2000.0),
with a 2$\sigma$ uncertainty of 0\fdg11. This position, consistent with
that obtained by \citet{joh+13} within the uncertainties, is $\sim$0\fdg11
away from the center of NGC~6626 while within the GC's tidal radius.
\citet{joh+13} discussed the possible origin for the emission. Since our purpose
is to constrain the number of other MSPs in the GC, we performed the binned
likelihood analysis to the off-pulse data with the position used.
The results (where TS$\simeq 139$) are given in Table~\ref{tab:phar}, and 
it can be noted that
the cutoff energy $E_c$ was poorly constrained. We tested to use a 
PL model in the analysis, and obtained a TS value of $\simeq$132
(photon index $\Gamma\simeq2.35\pm0.08$). Comparing the results from the
PLEC and PL model, 
the spectral curvature of the emission had a significance of only 
$\sim$2.6$\sigma$. In any case given our study purpose and the source spectrum 
(below we extracted), we adopted the results from the PLEC model.
\begin{figure*}
\centering
\includegraphics[angle=0,scale=0.6]{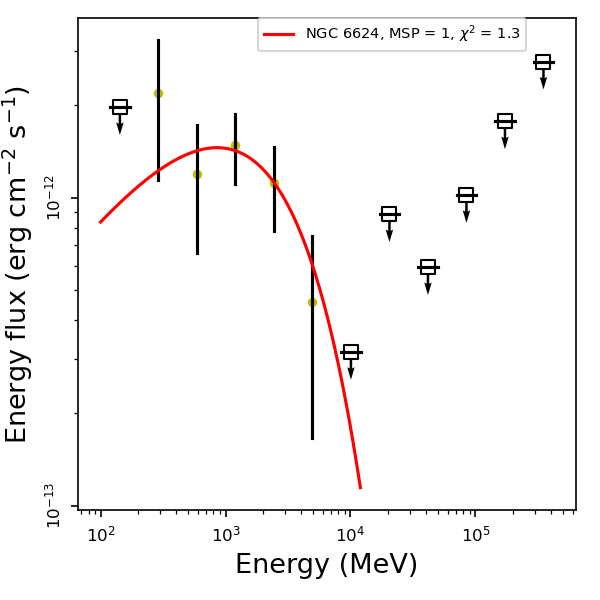}
\includegraphics[angle=0,scale=0.6]{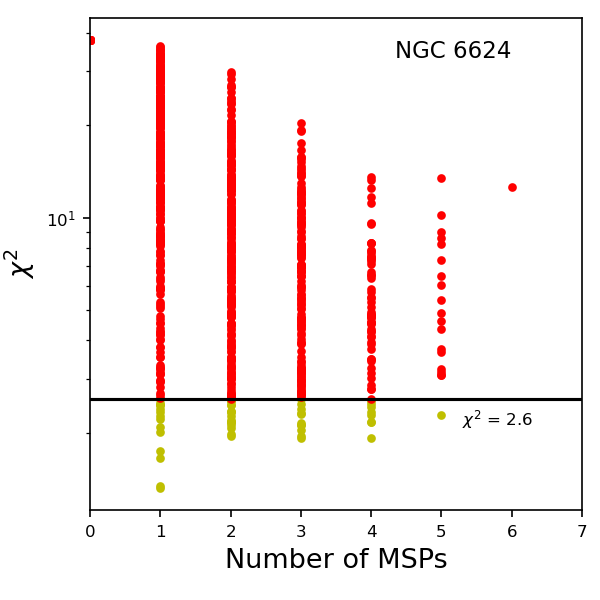}
\includegraphics[angle=0,scale=0.6]{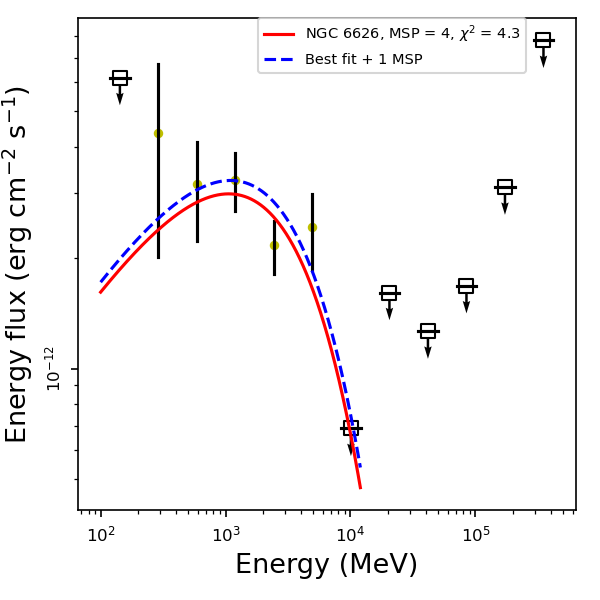}
\includegraphics[angle=0,scale=0.6]{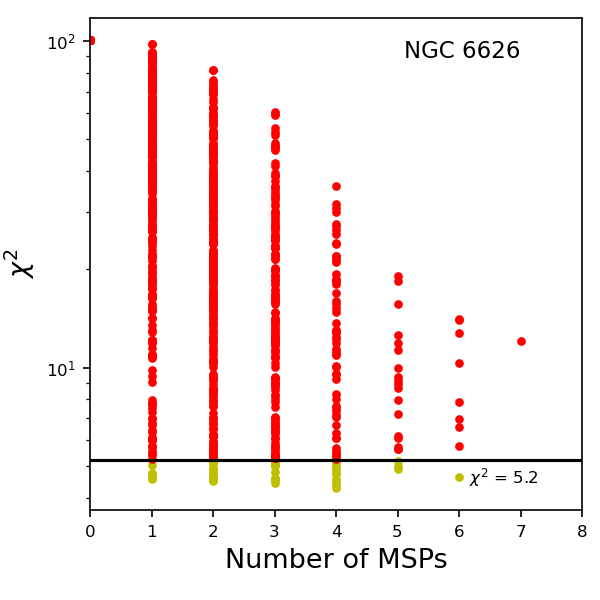}
	\caption{{\it Left} panels: off-pulse \gr\ spectra of PSR~J1823$-$3021A 
	(upper) and B1821$-$24 (lower). The best-fit models, which take
	the number of MSPs into account, are shown as red curves. In the lower
	panel, an example is shown when the addition of one more MSP causes 
	the model spectrum (blue dashed curve) greater than an upper-limit
	data point.
	{\it Right} panels: $\chi^2$ values from 1000 runs of spectral fitting,
	among which 
	the 5\% smallest values are marked as golden data points. 
	The 5\% limits (black lines) give a range of 1--5 and 1--6 
	respectively for the numbers of MSPs in NGC~6624 (upper) and NGC~6626
	(lower). }
	\label{fig:2426}
\end{figure*}

Pulse-phase resolved analysis for the emission of NGC~6652 was conducted
by \citet{zxw22}, and in a small off-pulse phase range they defined, no 
emission was detected (TS$\simeq$1). Using the results, we calculated 
a TS map in 0.1--500\,GeV and include it in Figure~\ref{fig:offtsmaps}.
The TS map confirms the non-detection of any emission in the off-pulse phase
range.

\begin{figure*}
	\centering
\includegraphics[angle=0,scale=0.6]{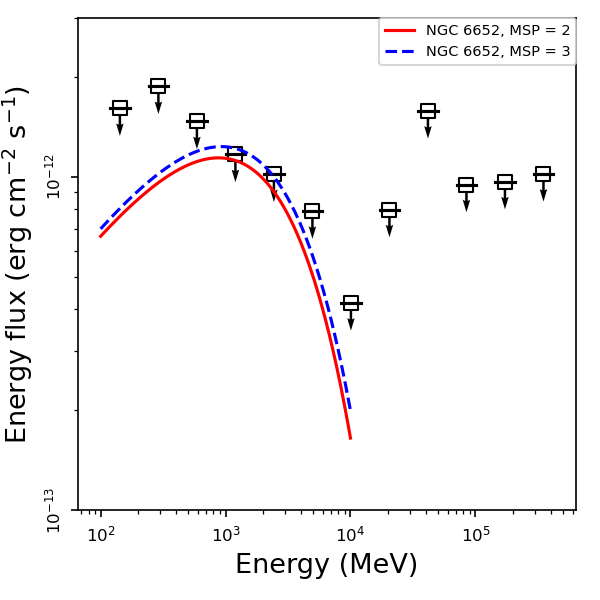}
\includegraphics[angle=0,scale=0.6]{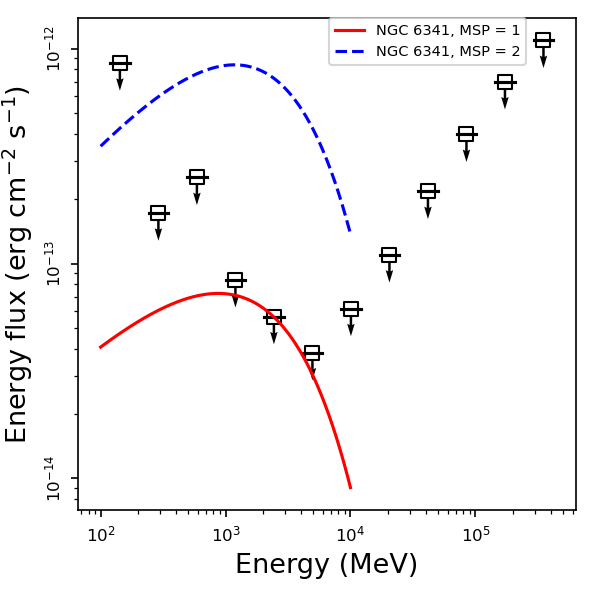}
	\caption{Constraints on the numbers of MSPs in NGC~6652 ({\it left})
	and NGC~6341 ({\it right}) from their spectral upper limits. The blue
	dashed line in each panel shows an example when the model spectrum 
	is greater than some upper-limit data points; the constraints on
	the numbers of MSPs are thus obtained.}
\label{fig:5241}
\end{figure*}

We extracted spectra of NGC~6624 and NGC~6626 from the whole data and
those from the data in the on-pulse and off-pulse phase ranges of
PSRs~J1823$-$3021A and B1821$-$24 respectively. For NGC~6652, we obtained its
spectral upper limits in the off-pulse phase range.
The energy range of 0.1--500\,GeV was divided into 12 equal logarithmically
spaced energy bins.
The maximum likelihood analysis was performed to the data in each energy bin.
The obtained spectra of NGC~6624 and NGC~6626, as well as 
the on-pulse ones, are shown in Figure~\ref{fig:spec2}, and those off-pulse
ones of NGC 6624 and NGC~6626 and spectral upper limits of
NGC~6652 are shown in Figures~\ref{fig:2426} \& \ref{fig:5241}. 
For the spectra, we only kept the flux data points with TS$\geq$4,
and the otherwise upper limits were at a 95\% confidence level.


\section{Discussion and Summary}
\label{sec:dis}
\subsection{PSR J1717$+$4308A in the GC NGC~6341}

Given the discovery and timing results for the MSP J1717$+$4308A 
in the GC NGC~6341 reported by \citet{pan+20,pan+21},
we have conducted detailed analysis of the \emph{Fermi}-LAT data.
A new \gr\ source close to the GC has been found. With this new source 
included in our analysis, we have studied the emission of the GC and
found that it can be
fitted with the PLEC model, one typically used for describing pulsars' 
emission. A $4.4\sigma$ pulsational signal has been detected at 
PSR~J1717$+$4308's spin period, although the H-test curve indicates that 
the early data before approximately year 2014 did not contribute to 
the significance of the signal.
This problem is likely caused by that the ephemeris of 
the MSP has been established in recent radio observations. When 
more \fermi-LAT data are collected in the near future, the significance 
would be expected to increase accordingly and the detection of the pulsational
signal will be more highly confirmed. 

Based on the pulse profile of the MSP we have obtained, phase-resolved 
analysis of the \gr\ data has been conducted.
No significant emission has been detected in the off-pulse phase range, 
and the TS maps calculated from the on-pulse and off-pulse phase ranges
support the detection of the pulsations of the MSP. 

\begin{figure*}
\centering
\includegraphics[angle=0,scale=0.3]{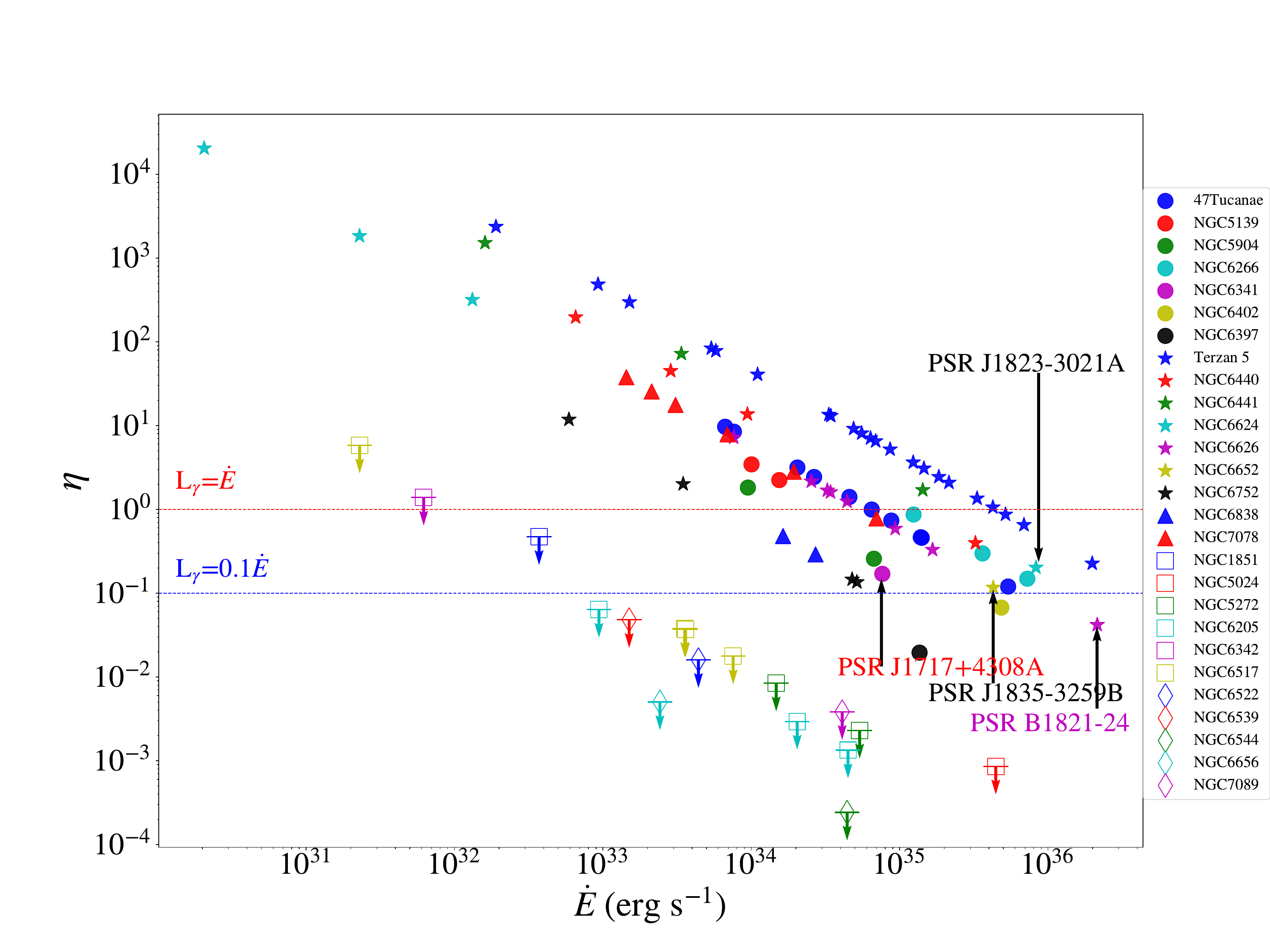}
	\caption{\gr\ efficiencies of PSRs J1823$-$3021A, B1821$-$24, 
	J1835$-$3259B, and J1717+4308A. Other known radio pulsars in the GCs 
	are included; for those in the GCs with \gr\ emissions, 
	the \gr\ luminosities of the GCs
	(but for NGC~6624 and NGC~6626, the \gr\ luminosities
	in the $\rm P_{off}$ phase ranges given in Table~\ref{tab:phar}) are
	used, and for those in the GCs without \gr\ emissions (marked by 
	downward arrows), a flux upper limit of 
	$10^{-12}$\,erg\,s$^{-1}$\,cm$^{-2}$ is used. The \gr\ luminosities
	and flux upper limit are derived and given in \citet{wu+22}.}
	\label{fig:emsp}
\end{figure*}

Considering that there is negligible emission in the off-pulse phase range,
the phase-averaged luminosity of the MSP
is $L_{\gamma}\simeq 1.3\times10^{34}$\,erg\,s$^{-1}$. The derived
spin-down energy of the MSP is 
$\dot{E}=7.7\times 10^{34}$\,erg\,s$^{-1}$, giving a \gr\ efficiency
$\eta$ of 0.17. This efficiency value is typical for \gr\ MSPs (e.g.,
\citealt{wu+22}). In Figure~\ref{fig:emsp}, we show this pulsar's value along
with the other three known GC \gr\ MSPs, in which the luminosities in the 
$\rm P_{on}$ phase ranges of PSRs J1823$-$3021A and B1821$-$24 
(Table~\ref{tab:phar}) and the phase-averaged 
luminosity of PSR~J1835$-$3259B (given no significant off-pulse emission
from this pulsar)
are used. In addition in the figure, we also show all the other radio pulsars 
in 27 GCs whose $\dot{E}$ values can be calculated, i.e., positive 
$\dot{P}$ values given in the GC pulsar table (note that the observed $\dot{P}$
values are mostly caused by dynamical effects and thus the $\dot{E}$ values
are rough estimates; see, e.g., \citealt{fre+17} and references therein). 
Among the GCs, 16 have 
reported $\gamma$-ray emissions and the other 11 do not. To plot the radio
pulsars in the figure, we assume 1) the off-pulse \gr\ luminosities for each
of NGC~6624 and NGC~6626 (Table~\ref{tab:phar}) as the luminosities for 
other known radio pulsars respectively in the two GCs 
(\citealt{abb+22}; \citealt{dou+22}; and references therein), 
and 2) the \gr\ 
luminosity of each 12 GCs (in which no \gr\ pulsars are identified) as 
that of each of
their radio pulsars (where the values are from \citealt{wu+22}). 
In NGC~6652, PSR~J1835$-$3259B is the only pulsar with $\dot{E}$ 
\citep{dec+2015, gau+2022}.
As can be seen, while the previous three MSPs are among the ones with the
highest $\dot{E}$, PSR~J1717+4308A has comparably low $\dot{E}$. One common
feature may be noted for the four \gr\ pulsars are that they are either supposed
to be bright given high $\dot{E}$ (cf., PSR~B1821$-$24), or their
host GCs are observed at radio or predicted at $\gamma$-rays \citep{wu+22}
to contain only several pulsars (i.e., possibly less contamination from 
other pulsars when searching for pulsational signals). 
This feature could be the reason for 
the detectability of the \gr\ pulsations of these four pulsars. 

\subsection{Implication for \gr\ emissions of GC MSPs}

With the likely \gr\ detection of PSR J1717+4308A, we take these four MSPs 
as the representative cases and investigate the implication for the MSPs that
would exist in the four host GCs. We use the off-pulse spectra of NGC~6624 and
NGC~6626 and the spectral upper limits for NGC~6652 and NGC~6341 (since the
latter two have not been found with significant off-pulse \gr\ emissions) and
perform the fitting to the spectra or upper limits. The method and related
MSP samples used have been fully described in \citet{wu+22} and 
the parameters for
the four GCs are given in Table~\ref{tab:four}. From fitting the off-pulse
spectra, the numbers
of other MSPs in NGC~6624 and NGC~6626 are constrained to be 1--5 and 1--6 
respectively (Figure~\ref{fig:2426}). The numbers well match those of the known 
radio pulsars (cf., Figure~\ref{fig:emsp}; \citealt{abb+22,dou+22}; and references therein), 
while it should be noted that several pulsars
in the two GCs do not have the spin-down rate information\footnote{https://www.naic.edu/$\sim$pfreire/GCpsr.html} (i.e., not shown in Figure~\ref{fig:emsp}).
Further, there is a caveat that the off-pulse emissions are assumed 
not containing any of the two MSPs,
which may not be true and possibly lead to over-estimates for the
numbers of other MSPs in the two GCs.
The numbers of other MSPs in NGC~6652 and NGC~6341 are constrained to be
$\leq$2 and $\leq$1 respectively. The constraints match the numbers
of known pulsars in the two GCs also well as they have thus-far been reported
to contain 2 and 1
pulsars respectively (including PSR J1835$-$3259B and PSR J1717+4308A; \citealt{gau+2022,pan+20}).
\begin{table}
	\centering
	\caption{Numbers of other \gr\ MSPs ($N^{\gamma}_{\rm MSP}$) estimated 
	from the off-pulse spectra or upper limits for the four GCs} 
                        \label{tab:four}
                        \begin{tabular}{lccccccc}
                                \hline
				Name & Age$^\ast$ & Dist$^\dagger$ & $\chi^{2}$/N & $N^{\gamma}_{\rm MSP}$ & $N^{5\%}_{\rm MSP}$\\
                                & (Gyr) & (kpc) &   & \\
                                \hline
                                  NGC 6624   & 11.7   & 8.0    & 2.6/5 & 1   & 1--5 \\
                                  NGC 6626   & 14.0   & 5.4    & 4.3/5 & 4   & 1--6 \\
				  NGC 6652   & 11.3   & 9.5    &       & $\leq$2   &        \\
                                  NGC 6341   & 12.8   & 8.5    &       & $\leq$1   &        \\
                                  \hline
                        \end{tabular}
		{Notes. $N^{5\%}_{\rm MSP}$ provides a range given by 5\% 
		lowest $\chi^2$ values in 1000 fitting runs (see \citealt{wu+22} for
		details). $^\ast$Ages of NGC~6624 and NGC~6626 are from \citet{oli+20}
		and \citet{ker+18} respectively, and those of NGC~6652 and NGC~6341 from \citet{van+13}. $^\dagger$ Distances are from \citet{bv21}.}
                \end{table}

Thus for the four GCs, the estimated MSP numbers roughly match the observed 
ones (which may suggest that the detectable pulsars 
in them have likely been found completely from radio and \gr\ surveys).
The results help
indicate that while the emission from one bright MSP can be dominant
in the observed \gr\ emission of the host GC, the contribution from other 
MSPs (at least for NGC~6624 and NGC~6626) is not negligible. 
The estimation for the number of \gr\ MSPs
in a GC should be carried out with caution, probably on a case-by-case basis
by taking into account the information provided from radio observations.
We have also learned from the four GCs that the presence of one bright MSP can
be the dominant factor for whether a GC has detectable \gr\ emission or not.

We might extend the thinking to all the GCs, that is the pulsars found in 
them are close to be complete, given the particular effort that 
has been made towards finding GC pulsars in the past with different facilities
(e.g., \citealt{pan+21, abb+22, dou+22, gau+2022}).
Then according to Figure~\ref{fig:emsp}, for which a companion table 
(Table~\ref{tab:gcmsp}) is made to provide more detailed information,
we should make an effort to search 
for \gr\ MSPs in Terzan~5, NGC~6266, NGC~6440, 47~Tuc, and probably NGC~5904 and
NGC~6752 as well; the latter two contain known MSPs with $\dot{E}$ similar
to that of PSR~J1717+4308A. Here we consider those pulsars falling
in the range of $\eta \sim$0.1--1. Also presented in Figure~\ref{fig:emsp}
(as well as in Table~\ref{tab:gcmsp})
are 11 GCs containing known radio pulsars but not having detectable 
\gr\ emissions, where
a flux upper limit of $10^{-12}$\,erg\,cm$^{-2}$\,s$^{-1}$ is used (estimated
from the work in \citealt{wu+22}). 
It can be seen that there is one MSP, PSR~J1312+1810E,
in NGC 5024 with $\dot{E} >$10$^{35}$\,erg\,s$^{-1}$. No emission has been
found from this GC (e.g., \citealt{yua+22}), while it can be noted that
the distance to the GC is large, $\sim$18 kpc \citep{Harris1996,bv21},
which may explain the non-detection at $\gamma$-rays of the GC (or the MSP).
In any case, this MSP is of interest as its $\eta$ already reaches 
$\sim10^{-3}$.


\subsection{Summary}

We have detected the \gr\ pulsations of PSR J1717+4308A in the GC NGC 6341
(M92) at a 4.4$\sigma$ confidence level. No significant off-pulse emission
from the MSP has been found, which suggests that the MSP's emission likely 
contributes
dominantly to the observed \gr\ emission of the GC. The detection has added
the fourth one to the group of GC \gr\ MSPs. Based on the four cases,
we may suggest that the
detectability of a \gr\ MSP in a GC relies on sufficiently high $\dot{E}$ 
(maybe greater than $10^{35}$\,erg\,s$^{-1}$) and
probably low number of other pulsars as well.

We have re-analyzed the \fermi-LAT data for the three previously known GC \gr\
MSPs and obtained their off-pulse spectra and spectral upper limits. Fitting
the spectra or upper limits with the method used in \citet{wu+22}, which
include the spectral upper limits on NGC~6341, we have
constrained the numbers of other MSPs respectively in the four GCs.
The numbers are in consistency with those of known radio
pulsars in the four GCs, suggesting that probably most of pulsars in them 
have already been found. While at least in NGC~6624 and NGC~6626, the
\gr\ emission from other pulsars is not negligible, the four GCs show that
a bright MSP can be a dominant factor for whether a GC is detectable at
$\gamma$-rays or not. Our study has also suggested a few targets for finding 
more GC \gr\ MSPs.
\begin{table*}
\begin{center}
	\caption{Detailed information for the estimated numbers of \gr\ MSPs 
	and the numbers of reported radio MSPs in 23 GCs shown in Figure~\ref{fig:emsp}}
\begin{tabular}{cccccccc}
\hline\hline
	\multirow{2}{*}{GC name} & $L_\gamma/10^{34}$  & \multirow{2}{*}{$N^{\gamma}_{\rm MSP}$} &  \multirow{2}{*}{$N_{\rm MSP}$} & \multirow{2}{*}{MSP name} & $P$ & $\dot{P}/10^{-19}$ & $\dot{E}/10^{34}$ \\
	& (erg s$^{-1}$) & & & & (ms) & (s s$^{-1}$) & (erg s$^{-1}$)\\
\hline
47 Tucanae &6.5 & 1--11 & 29 & J0024--7205E & 3.54 & 0.99 & 8.8\\
                    &        &         &      & J0024$-$7204F & 2.62 & 0.65 & 14.1\\
                    &        &         &      & J0024$-$7204O & 2.64 & 0.30 & 6.5\\
                    &        &         &      & J0024$-$7204R & 3.48 & 1.48 & 13.9\\
\hline
NGC 5139   & 3.5 & 1--6   & 18 & --- & & &  \\
\hline
NGC 5904   &1.7 & 1--3   & 7   & B1516+0204C & 2.48 & 0.26 & 6.7\\
\hline
NGC 6266 & 10.8& 1--13 & 9 & J1701$-$3006D & 3.42 & 1.26 & 12.4\\
                  &         &          &    & J1701$-$3006E & 3.23 & 3.10 & 36.2\\
                  &         &          &    & J1701$-$3006F & 2.29 & 2.22 & 72.6\\
\hline
NGC 6402 & 3.3   & 1--6   & 5 & J1737$-$0314A & 1.98 & 0.95 & 48.5\\
\hline
NGC 6397 & 0.3   & 1--1   & 2 & J1740$-$5340(A)& 3.65 &1.68 & 13.6\\
\hline
Terzan 5 & 45.0 & 9--38 & 41 & J1748$-$2446P & 1.73 & 2.60 & 199.0\\
               &         &          &      & J1748$-$2446Y & 2.05 & 1.50 & 68.9\\
               &         &          &      & J1748$-$2446ak&1.89 & 0.88 & 51.7\\
\hline
NGC 6440 & 12.9 & 1--11 & 7 & J1748$-$2021H & 2.85 & 1.90 & 32.5\\
\hline
NGC 6441 & 24.4 & 4--20   & 7 & --- & & &  \\
\hline
NGC 6752 & 0.7 & 1--2 & 9 & J1910$-$5959D & 9.04 & 9.63 & 5.2\\
         &     &      &   & J1910$-$5959F & 8.49 & 7.41 & 4.8\\
\hline
NGC 6838 & 0.8 & 1--2 & 3 & J1953+1846A & 4.89 & 0.49 & 1.64\\
                 &          &        &    & J1953+1846E & 4.44 & 0.60 & 2.72\\
\hline
NGC 7078 & 5.5  &  1--5  & 5 & B2127+11E     & 4.65 & 1.78 & 6.98\\
\hline
NGC 1851 & 0.02$^{\rm a}$ & ---$^{\rm b}$ & 14 & J0514$-$4002A & 4.99 & 0.01 & 0.04\\
\hline
NGC 5024 & 0.04$^{\rm a}$ & ---$^{\rm b}$ & 4 & J1312+1810E & 3.97 & 7.05 & 44.5\\
\hline
NGC 5272 & 0.01$^{\rm a}$ & ---$^{\rm b}$ & 6 & J1342+2822B & 2.39 & 0.19 & 5.4\\
                  &           &      &    & J1342+2822F & 4.40 & 0.32 & 1.5\\
\hline
NGC 6205 & 0.006$^{\rm a}$ & ---$^{\rm b}$ & 6 & J1641+3627C & 3.72 & 0.01 & 0.1\\
                  &           &   &    & J1641+3627E & 2.49 & 0.18 & 4.5\\
                  &           &  &    & J1641+2627F & 3.00 & 0.14 & 2.0\\
\hline
NGC 6342 & 0.009$^{\rm a}$ &  ---$^{\rm b}$   & 1 & --- & & &   \\
\hline
NGC 6517 & 0.01$^{\rm a}$ & ---$^{\rm b}$ & 16 & J1801$-$0857B & 28.96 & 21.91 & 0.4\\
                  &          &     &      & J1801$-$0857D & 4.22 & 0.06 & 0.4\\
                  &          &     &      & J1801$-$0857H & 5.64 & 0.34 & 0.8\\
\hline
NGC 6522 & 0.007$^{\rm a}$ & ---$^{\rm b}$ & 5 & J1803$-$3002A & 7.10 & 0.40 & 0.4\\
\hline
NGC 6539 & 0.007$^{\rm a}$ & ---$^{\rm b}$ & 1 & B1802$-$07 & 23.10 & 4.70 & 0.2\\
\hline
NGC 6544 & 0.001$^{\rm a}$ & ---$^{\rm b}$ & 3 & J1807$-$2500B & 4.19 & 0.82 & 4.4\\
\hline
NGC 6656 & 0.001$^{\rm a}$ & 1--1 & 4 & J1836$-$2354A & 3.35 & 0.02 & 0.2\\
\hline
NGC 7089 & 0.015$^{\rm a}$ & ---$^{\rm b}$ & 6 & J2133$-$0049D & 4.22 & 0.78 & 4.1\\
\hline\hline
\end{tabular}
\label{tab:gcmsp}
\end{center}
    {Notes. $N^{\gamma}_{\rm MSP}$ is the number of \gr\ MSPs in a GC 
    estimated in \citet{wu+22}, and $N_{\rm MSP}$ is the number of MSPs
    ($P\leq 30$\,ms) listed in the GC pulsar table, while the MSPs with 
    \gr\ efficiency $\eta\leq 1$ (cf., Figure~\ref{fig:emsp}) are specifically
    listed as potential \gr\ sources.  $^{\rm a}$Luminosity upper limits,
    where for the GCs with no detected $\gamma$-rays, a flux upper limit of
    $10^{-12}$\,erg\,s$^{-1}$\,cm$^{-2}$ is used. $^{\rm b}$ GCs not appearing 
    in \citet{wu+22}. }
\end{table*}

\begin{acknowledgments}
	We thank the anonymous referee for helpful comments.
This work is supported in part by the National Natural Science Foundation of 
China No.~12163006 and 12233006, the Basic Research Program of Yunnan Province No.~202201AT070137,
and the joint foundation of Department of Science and Technology of
Yunnan Province and Yunnan University (202201BF070001-020).
Z.~W. acknowledges the support by the Basic Research Program of Yunnan 
	Province (No. 202201AS070005), the National Natural Science Foundation 
	of China (12273033), and the Original
Innovation Program of the Chinese Academy of Sciences (E085021002).

\end{acknowledgments}

\bibliographystyle{aasjournal}
\bibliography{aas}
\end{document}